\documentclass{emulateapj}

\shorttitle{QPPs in the 2009 July 5 Solar Flare}
\shortauthors{Rao et al.}

\begin{document}
\title{RT-2 Detection of Quasi-Periodic Pulsations in the 2009 July 5 Solar
Hard X-ray Flare}

\author{A. R. Rao\altaffilmark{1}, J. P. Malkar\altaffilmark{1}, 
M. K. Hingar\altaffilmark{1}, V. K. Agrawal\altaffilmark{1,2}, 
S. K. Chakrabarti\altaffilmark{3,4}, A. Nandi\altaffilmark{4,2}, 
D. Debnath\altaffilmark{4}, T. B. Kotoch\altaffilmark{4}, 
T. R. Chidambaram\altaffilmark{5}, P. Vinod\altaffilmark{5}, 
S. Sreekumar\altaffilmark{5}, Y. D. Kotov\altaffilmark{6}, 
A. S. Buslov\altaffilmark{6}, V. N. Yurov\altaffilmark{6}, 
V. G. Tyshkevich\altaffilmark{6}, A. I. Arkhangelskij\altaffilmark{6}, 
R. A. Zyatkov\altaffilmark{6}, S. Shaheda Begum\altaffilmark{7}, 
P. K. Manoharan\altaffilmark{7}}

\altaffiltext{1}{Tata Institute of Fundamental Research, Mumbai - 400005, 
India. $arrao@tifr.res.in$} 
\altaffiltext{2}{Space Science Division, ISRO Head Quarters, New Bel Road, Bangalore - 560231,
India.}
\altaffiltext{3}{S. N. Bose National Centre for Basic Sciences, Salt Lake, Kolkata - 700098,
India.}
\altaffiltext{4}{Indian Center for Space Physics, 43 Chalantika, Garia St. Rd., Kolkata 
- 700084, India}
\altaffiltext{5}{Vikram Sarabhai Space Centre, VRC, Thiruvananthapuram - 695022, India.}
\altaffiltext{6}{Moscow Engineering Physics Institute, Moscow - 115409, Russia.}
\altaffiltext{7}{Radio Astronomy Center, Ooty - 643001, India.}

\begin{abstract}
We present the results of an analysis of hard X-ray observations
of the C2.7 solar flare detected by the $RT$-2 Experiment onboard
the $Coronas-Photon$ satellite. We detect hard X-ray pulsations 
at periods of $\sim$12 s and  $\sim$15 s.  We find a marginal
evidence for a decrease in period with time. We have augmented these results using 
the publicly available data from the $RHESSI$ satellite. We present
a spectral analysis and measure the spectral parameters. 
\end{abstract}

    
\section{Introduction}

  Quasi-Periodic Pulsations (QPPs), a common feature of solar flare emission,
have been observed for many years (\citet{Young61}) in all frequency
bands ranging from radio to hard X-rays with periodicities varying from  a
few milliseconds to several seconds (\citet{Aschwanden87}; \citet{Fleishman02}; 
\citet{Tan08}). The long period QPPs (periodicity $>$ 10 s)
observed in the microwave emission of solar flares are also seen in
hard X-rays (\citet{Kane83}; \citet{Nakajima83}; \citet{Asai01}; 
\citet{Nakariakov03}) and they could be
resulting from some MHD oscillations in the source region or due to 
modulation of electron acceleration and injection mechanisms.
QPPs are generally associated with the emission from the flare 
accelerated non-thermal 
electrons, because thermal parameters are not expected to
show sudden changes
and pulsations. QPPs can be observed in all the stages of  a  flare,
prominently in hard X-rays, microwave and white light emissions. 

MHD oscillations in the magnetic loop can cause modulation in the magnetic
field strength and magnetic mirror ratio (\citet{Zaitsev82}, \citet{Zaitsev89}; 
\citet{Zimovets09})
resulting in periodic variation towards the flare foot points. 
Generally, most of the non-thermal hard X-rays are observed at the
foot points of the magnetic loops involved in the magnetic
reconnection. This emission is produced by the accelerated non-thermal
electrons from the reconnection region hitting the foot points of the magnetic
loops. The oscillations in the accelerated electrons in turn reflect as QPPs
in the hard X-ray emission profile.
   Similarly, microwave emission produced by the interaction of the 
accelerated electrons with the magnetic field as a result of gyro-synchrotron
process also show QPPs (like hard X-rays, as both the emissions are from the 
same population of electrons).
   Observationally, QPPs can be seen in the light curves of the 
solar flare emission in the respective wave-bands. 
The profiles of  photons or energy fluxes 
of hard X-rays associated with  the 
non-thermal electrons show QPPs or damping oscillations.

 Since the basic cause of QPPs have implications for particle acceleration 
mechanism,
it is important to investigate QPPs at diverse source intensities.
In this paper, we present the results obtained from the observation
of the C2.7 solar flare detected on 2009 July 5 using  the RT-2
experiment onboard the Coronas-Photon satellite. Since this is the 
first result from this experiment, we describe in detail the methodology
used in deriving the response matrix and spectral fitting. We augment
our results by using the publicly available $RHESSI$  data. 
We examine the spectral and temporal characteristics
of the flare and investigate the implications to the electron acceleration
mechanisms. 
In \S 2 a brief description of the RT-2 experiment is given. Observations
and analysis results (RT-2 and RHESSI data) are given in \S 3 and finally
in \S 4 a detailed discussion of the results are presented along
with relevant conclusions.

\section{RT-2 Experiment onboard Coronas-Photon satellite}

RT-2 Experiment (RT - Roentgen Telescope), is a part of 
the Coronas-Photon mission, launched on 2009 January 30 (\citet{Kotov08}; 
\citet{Nandi09a}). 
The primary objective of the mission is to make a detailed
temporal and spectral study of hard X-ray and gamma-ray 
emission during solar  flares. The satellite is in a near polar
(inclination 82$^{\circ}$.5), near-earth (altitude 550 km)
Sun-synchronous orbit. Though the large inclination gives low
duty cycle of observation due to the increased
background emission at high latitudes and South Atlantic Anomaly (SAA)
regions, it facilitates Sun synchronization for long uninterrupted
observations of the Sun. 

RT-2 consists of an ensemble of the low energy gamma-ray (or  hard X-ray) 
detectors sensitive in the energy range of 15 keV to 150 keV and
also having an extended detection capability up to 1000 keV.
It consists of three instruments called 
RT-2/S,  RT-2/G, and RT-2/CZT.
RT-2/S and RT-2/G detector assemblies have an identical
configuration 
of NaI(Tl) / CsI(Na) scintillators in phoswich combination.
Both the detector assemblies  sit behind respective mechanical slat 
collimators surrounded by uniform shields of Tantalum material and having 
different viewing angles of 4$^{\circ}\times4^{\circ}$ (RT-2/S) and 
6$^{\circ}\times6^{\circ}$ (RT-2/G). The collimation is effective up to 
about 150 keV and above this energy these detectors act as omni-directional
low energy gamma-ray detectors with sensitivity up to $\sim$1000 keV.
The low background high sensitivity range for RT-2/S is 15 to 150 keV
whereas the use of an aluminum filter in RT-2/G 
sets the lower energy threshold at 25 keV. 
The RT-2/CZT consists of three CZT 
detector modules (OMS40G256) and one CMOS detector (RadEye-1) arranged in 
a 2$\times$2 array. 
The CZT-CMOS detector assembly is mounted 
behind a collimator with two different types of coding devices, namely 
coded aperture mask (CAM) and Fresnel zone plate (FZP), surrounded by a 
uniform shield of Tantalum material and has a viewing angle of 
ranging from 6$'$ to 6$^{\circ}$. The RT-2/CZT payload is the only imaging 
device in the Coronas-Photon mission to image the solar flares in hard 
X-rays in the energy range of 20 to 150 keV. 

During the `GOOD' regions (that is, outside the high background regions of Polar 
Caps and SAA), the RT-2/S and RT-2/G generally operate in the
Solar Quiet Mode (SQM) when count rates in eight channels (for each detector) are 
stored 
every second. The spectral data are stored every 100 s. The low energy 
spectra are stored separately for NaI(Tl) 
(15 -- 150 keV)
and NaI(Cs) (25 -- 215 keV) detectors based on the pulse shape 
along with the 
high energy spectrum in the energy range of 215 -- 1000 keV
(see \citet{Debnath09} for details). The onboard software
automatically stores the data in finer time resolution (0.1 s count rates and
10 s spectra) during the Solar Flare Mode (SFM), when the count rates 
exceed a pre-determined limit. RT-2/CZT operates only in SQM when 1 s count rates
and 100 s spectra and images are stored.
The test and evaluation results of this payload are described in
\citet{Nandi09b}, \citet{Debnath09}, \citet{Kotoch09}, \citet{Sarkar09}, 
and \citet{Sreekumar09}.

\section{Observation and Analysis}

   The flare of class C2.7 occurred near the center of the disk (S27W12)
in NOAA active region 11024 on 2009 July 5, which peaked at 07:12 UT. 
From the X-ray light curves derived from RT-2, GOES and RHESSI observations, 
it can be concluded that this flare is compact and impulsive in 
nature.
During the  solar flare the RT-2/S and RT-2/G were in the Solar Quiet Mode.
The count rates were too low to trigger the Solar Flare Mode.
Due to some operational constraints, the low energy threshold of CZT 
detectors were kept at $\sim$40 keV and hence
RT-2/CZT did not detect this flare.

  The observed count rates of the solar flare from the 
RT-2/S detector are given in Figure 1 for low energies (20 -- 35 keV),
high energies (35 -- 59 keV), along with the count rates above 215 keV,
which represents the high energy particle background rates. 
The bin size is one second and the time is given
in UT seconds on 2009 July 5. 
The satellite
was at high latitudes at the beginning of the observations
and the background rates slowly stabilized when the satellite
approached the low background equatorial region. Since the detectors use
the Phoswich technique, the changing background has negligible impact
on the $<$ 35 keV light curves. For example, the count rate during the
first 200 second of observation (in this energy band) is 36.6$\pm$0.4 
s$^{-1}$ and it is 36.4$\pm$0.4 s$^{-1}$  towards the end of observation.
During this time the background rate decreased from 576$\pm$2 s$^{-1}$ to 
235$\pm$1 s$^{-1}$. There is, however, some decreasing trend in the
$>$  35 keV count rates. Solar flare starting from 25800 s UT (07:10) is seen clearly
in both the energy channels.

 The RT-2/G detector has an Aluminum window to block X-rays below $\sim$25 keV
and hence it samples the high energy photons. The light curves of
the solar flare as measured by the RT-2/S and RT-2/G detectors
are shown  in two channels each in Figure 2.   The 
bin size is 1 second and T$_0$ is UT 07:08:50 on 2009 July 5.
Quasi-periodic pulsations are clearly seen in the light curve. We
define the rising phase of the flare as between 125 to 225 s (07:10:55 to
07:12:35) and the falling phase as between 225 s to 325 s (07:12:35
to 07:14:15). These regions are demarcated by vertical dashed
lines in the top panel of Figure 2.
They also correspond to the availability of the spectral data in the
Solar Quiet Mode.


  The light curves obtained from the  GOES 10 and RHESSI satellites
are shown in Figure 3 and they  show that 
the flare is an impulsive one, where the rising phase takes 
approximately 4 minutes from the onset to peak flux.
The GOES   light curves in two channels (1 -- 8  $\AA$ : 1.6 -- 12.4 keV and
 0.5 -- 4 $\AA$ : 3.1- 24.8 keV) are shown in the top panel of the
figure with a bin size of 1 minute. The RHESSI light curves in three
energy bands  (3 -- 6 keV, 6 -- 12 keV,  and 12 -- 25 keV)
are shown in the bottom panel of the figure with a bin size of 4 seconds.

\subsection{Timing Analysis}

We follow the method given in Fleishman et al. (2008) to find the
modulation power and the periodicities. If C(t) is the count rate
at time t, the normalized modulation is 
$$
S(t) = \frac {C(t) - <C(t)>}{<C(t)>}
$$
where $<C(t)>$ is the running average taken over a number of bins, 20 s in 
our case. The modulation power over a period of time is the average of
S$^2$(t) and the square root of modulation power is the modulation amplitude
(see \citet{Fleishman08}).

 In  Figure 4, we have plotted the normalized modulation along with the
count rates, with a bin size of 1 second. 
The top panel shows the count rates in the RT-2/S 20 -- 35 keV range
 and the
successive panels downwards show the normalized modulation for
RT-2/S 20 -- 35 keV, 35 -- 59 keV and similarly for RT-2/G, respectively.
  To estimate the errors in the modulation power, we have calculated the
modulation power in the light curves outside the flares, in batches of 100 s, and the rms 
variation in them are deemed as the error in the measured values. The background
subtracted modulation powers, for 4 second integration time, are given in Table 1,
for the rising and falling phase of the flare, respectively.
The modulation amplitudes are also calculated for the RHESSI data. 
The flare shows the highest modulation of 13.5\%in the low energy
RT-2/S band (20 -- 35 keV) and it is $\sim$5 -- 8\% in low energies
(from RHESSI data) as well as above 35 keV.

To investigate the values of periodicities and their variation, we have
followed the method used in \citet{Fleishman08} and calculated the
Fourier transform of the normalized modulation derived with 1 s time resolution.
We have shown the Fourier transform of the normalized modulation in Figure 5
for RT-2/S 20 -- 35 keV total light curve (top panel) and the rising phase of the 
flare (second panel from top). The third and fourth panel shows similar
power spectra for the RT-2/G light curves. 
To quantify the errors in periods and the significance level of
the period determination, we represent the amplitudes of the Fourier
components with a normalization as suggested by \citet{Horne86}. 
This  method has the
added advantage of quantifying the false alarm probability, that is the
probability of getting a power higher than the observed peak, by random
distribution. The highest peaks in the power
spectrum, along with the confidence levels (false alarm probability), 
are shown in Table 2.
 To increase the signal to noise ratio, we have added all
counts (two channels of RT-2/S and the lowest channel of RT-2/G) and
have given the corresponding peaks in Table 2.  
It can be seen from the table (also see Figure 5) that two prominent
peaks are seen in the periodogram corresponding to the periods of 
$\sim$12 s and $\sim$15 s, respectively. Both the periods are 
seen very significantly in the full data set, whereas the 15 s 
periodicity is more significant in the rising phase.

To investigate any possible period changes during the flare, we have used
a running window of 60 s duration and measured the periodicities in the
full data set. The variation of the period with time (starting from the
flare onset) is given in Figure 6. There is an indication 
of the value of period decreasing with time. A straight line fit to the
data gives a value of period derivative as -0.06$\pm$0.03 s/s for the
15 s periodicity and -0.02$\pm$0.01 s/s for the 12 s periodicity.

\subsection{Spectral Analysis}

 We have generated the appropriate response matrices of the RT-2/S and RT-2/G
detectors. Background lines at 56.9 keV (due to I$^{121}$
decay) are used for channel to energy conversion. The values
of energy resolution function measured during the ground calibration and
the effective areas from the known geometrical properties of the
detectors are  used for response matrix generation. Background spectrum
obtained away from the solar flare is used. The XSPEC tool of the $ftools$
package is used for spectral fitting. The deconvolved spectrum
is shown in Figure 7 for RT-2/S (filled circles) and RT-2/G (open
circles). The spectrum is very steep and it is best fit by
a simple bremsstrahlung function of energy 3.43$\pm$0.30 keV. 
This model is shown as 
a dashed line in the figure.

We have also generated
the spatially integrated, background subtracted spectra from RHESSI observations
for the rising phase and the falling phase  of the flare.
The count spectral files were created using the standard RHESSI 
software of Solar SoftWare (SSW). The data were accumulated over 30 seconds
with 97 energy bands from 3 to 100 keV using all front detector segments
excluding 2 and 7 (for their lower energy resolution and high threshold
energies respectively). The full spectrum response matrix was used to
calibrate the data. Then the RHESSI OSPEX package is used for spectral fitting
of the count spectra.

   The background spectra was accumulated for the non-flare period of (07:06:28 
to 07:06:58) 30 seconds in all the energy levels. The count spectra were fitted
with a two component model consisting of a optically thin thermal Bremsstrahlung radiation 
function
parameterized by the plasma temperature kT and the emission measure (EM) and
a thick target Bremsstrahlung characterized by the electron flux and
the power law index ($\Gamma$) of the electron distribution function below 
the break energy. 
The best fit
parameters, temperature kT of the isothermal emitting plasma and its emission 
measure EM are derived by these fitted spectra 
are given in Table 3. The deconvolved photon spectrum is given in Figure 8. 
It can be seen that the 20 -- 30 keV RHESSI spectrum agrees quite closely with the
RT-2 data.

\section{Discussion and conclusions}

\citet{Jakimiec09} have investigated QPPs in about
50 flares using $Yohkov$ and $BATSE$ hard X-ray data and have derived 
a correlation between the QPP periods (ranging from 10 s to 150 s) and
sizes of loop-top sources. From the RHESSI data we derive a size of 
the X-ray emitting region in the 6 -- 12 keV region of 7" (5 Mm), corresponding
to the 50\% level of the peak emission, for the 2009 July 5 flare. This size 
agrees with the correlation derived by \citet{Jakimiec09}.

\citet{Desai87} detected fast oscillations in several solar hard X-ray flares
and observed magneto-hydrodynamic signature of the loop dynamics.
\citet{Jakimiec09}
conclude that the hard X-ray 
oscillations are confined to the loop-top sources and the observations
are described with a model of oscillating magnetic traps. \citet{Fleishman08}
have made a detailed analysis of the 2003 June 15 solar flare (GOES X1.3 class)
and detected hard X-ray (based on $RHESSI$ data) and microwave oscillations
with periods ranging from 10s to 20 s. They, however, conclude that QPPs are
associated with quasi-periodic acceleration and injection of
electrons.
The possible detection of a decreasing trend in the periodicity can put 
further constraints on the magneto-hydrodynamic models.

Several flare observations as well as numerical simulation studies
 have been reported on the periodic and quasi-periodic oscillations of
 flare intensity in the radio and X-ray energy bands.
 Such oscillations show the typical size of reconnection site,
 configuration of loops formed during the reconnection,
 and plasmoid or CME launched above the reconnection X-point.

	As shown in Figure 2, the rising phase of the flare shows
 the nominal exponential increase. However, as the flare attains
 the peak level, the intensity goes through moderate quasi-periodic
 oscillations, which are more prominent in the 20 -- 35 keV energy band.
 Moreover, the modulation index (i.e., the degree of quasi-periodic oscillation)
 is higher at the low-energy band (i.e., $\sim$13.5\% in the 20 -- 35 keV band)
 than in the high-energy band ($\sim$7\% above 35 keV bands).
 It shows the production of copious amount of electrons over a limited
 range of energies. The flare profile observed at 15.4 GHz correlates
 with the rising phase of the flare, but the oscillations are not clearly
 seen, which may be a limitation imposed by the sensitivity of the measurement.

The white-light images from LASCO, associated with this flare, show
 a rather slow moving CME (i.e., speed in the range of 50 -- 150 km s$^{-1}$).
This is consistent with our finding of the production of particles in
a limited energy range.
 A comparison of profiles shown in Figure 2 with the RHESSI spectrum
 reveals a gradual steepening of the spectrum from the flare rising
 phase to the start of the decay phase, although the average intensity
 of the flare remained nearly same level in this period.
 Thus, most of the accelerated electrons have been generated and injected
 from the reconnection site.

 This is the brightest solar flare detected by the RT-2 experiment in
the first ten months of operation. Several other solar flares, particularly 
during the eruptions that have taken place from 2009 Oct 22 and Nov 2,
are also recorded. 
A detailed investigation on faint flares during this solar minimum are
going on and a flare list would be published separately. From 2009
December onwards, communication to the satellite is not responding, 
though attempts are on to revive the system.

\acknowledgements
This work was made possible in part from a grant from Indian Space Research Organization
(ISRO). The whole-hearted support from G. Madhavan Nair, Ex-Chairman, ISRO, who initiated 
the RT-2 project, is gratefully acknowledged. Significant contributions from several 
organizations for the realization of the RT-2 payload is gratefully acknowledged.
We are also grateful to the RHESSI and GOES spacecraft teams for the respective data.

\begin{figure*}
\includegraphics[angle=-90,scale=.70]{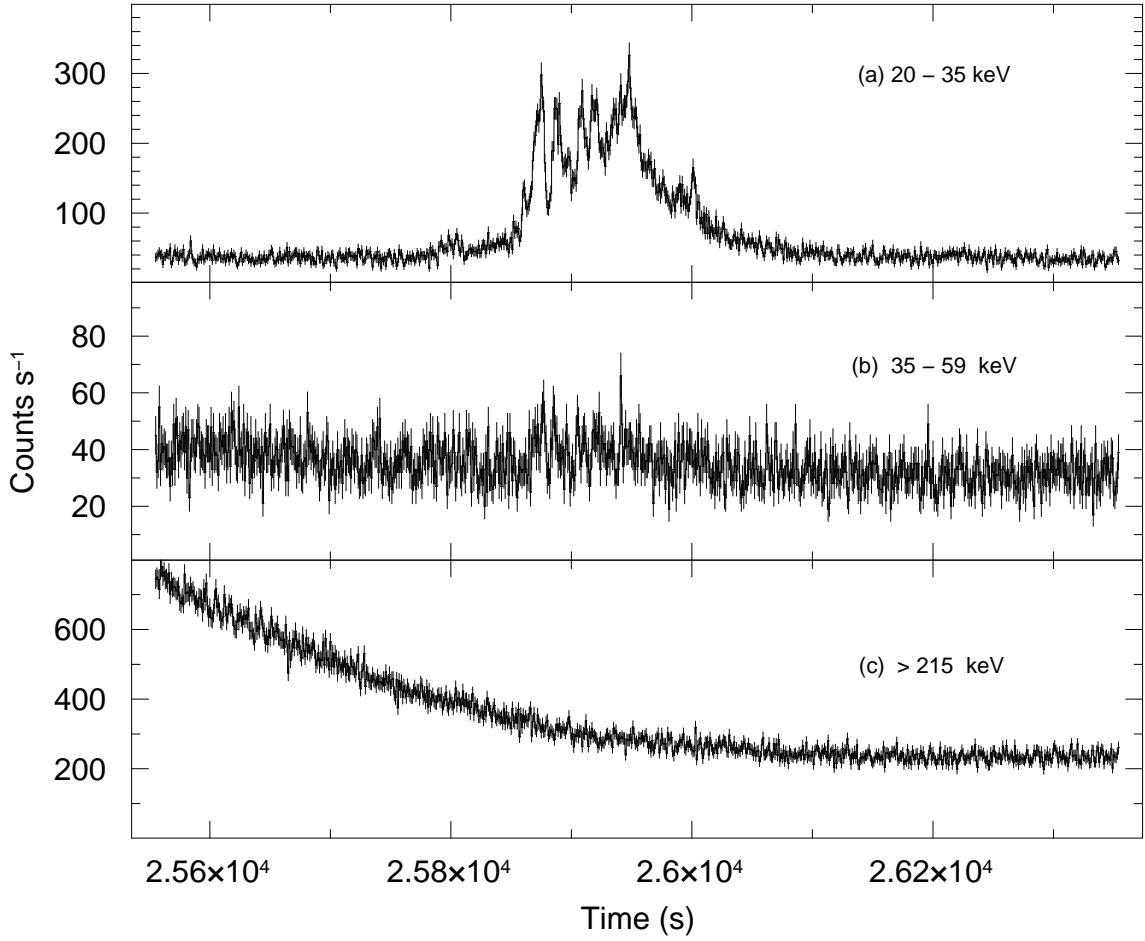}
\caption{The RT-2/S light curve of the solar flare in 20 -- 35 keV
(top panel) and 35 -- 59 keV (middle panel) 
with a bin size of 1 second. The background rate (above 215 keV)
is shown in the bottom panel. The time is in UT seconds on 2009
July 5.}
\end{figure*}

\begin{figure*}
\includegraphics[angle=-90,scale=.70]{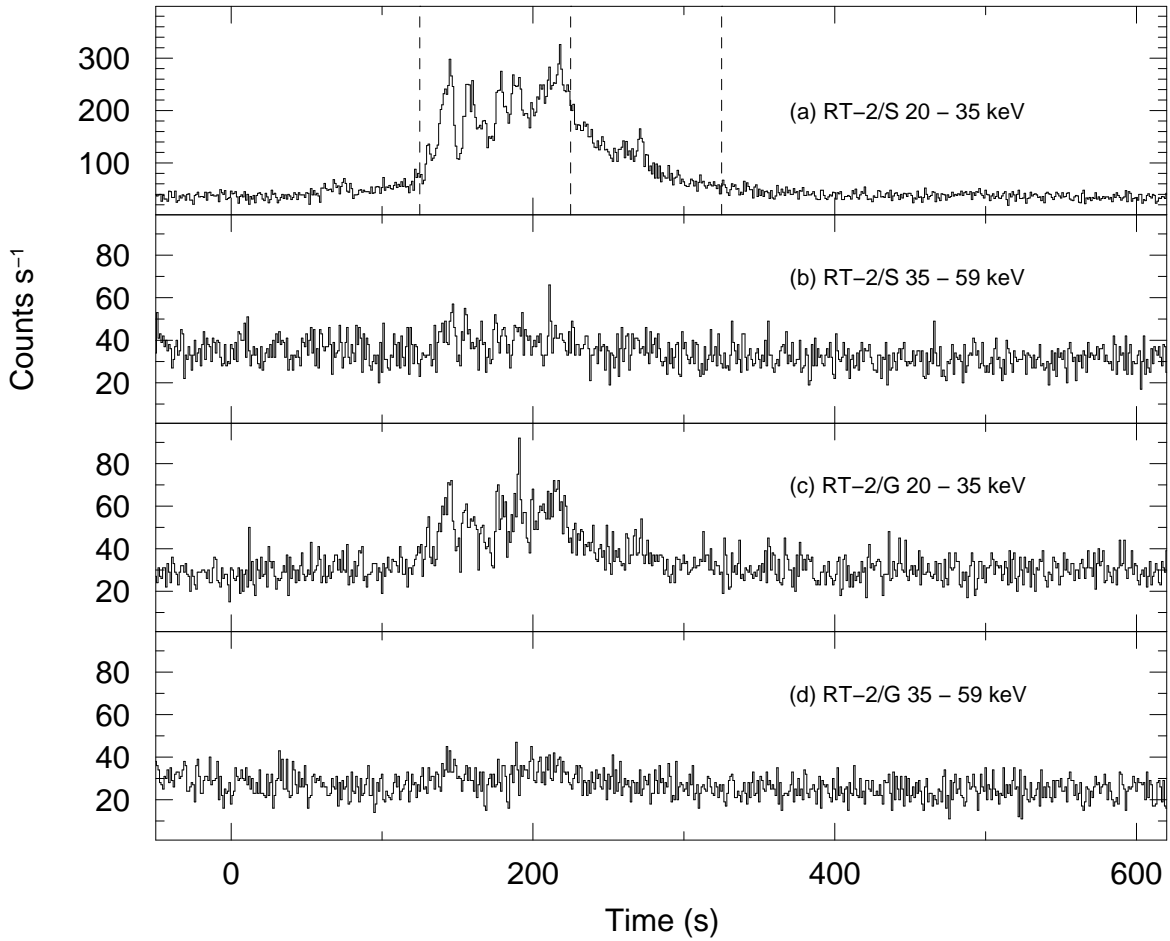}
\caption{The RT-2/S light curve in 20 -- 35 keV channel (top panel) and 35 -- 59 keV
channel (second from top) shown along with the  RT-2/G light curves 
(third and fourth panels). 
The bin size is 1 s and
T$_0$ is UT 07:08:50 on 2009 July 5. The vertical dashed lines in panel
(a) demarcate the rising (07:10:55 to 07:12:35) and falling (07:12:35 to
07:14:15) phase of the flare.}
\end{figure*}


\begin{figure*}
\includegraphics[angle=-90,scale=.70]{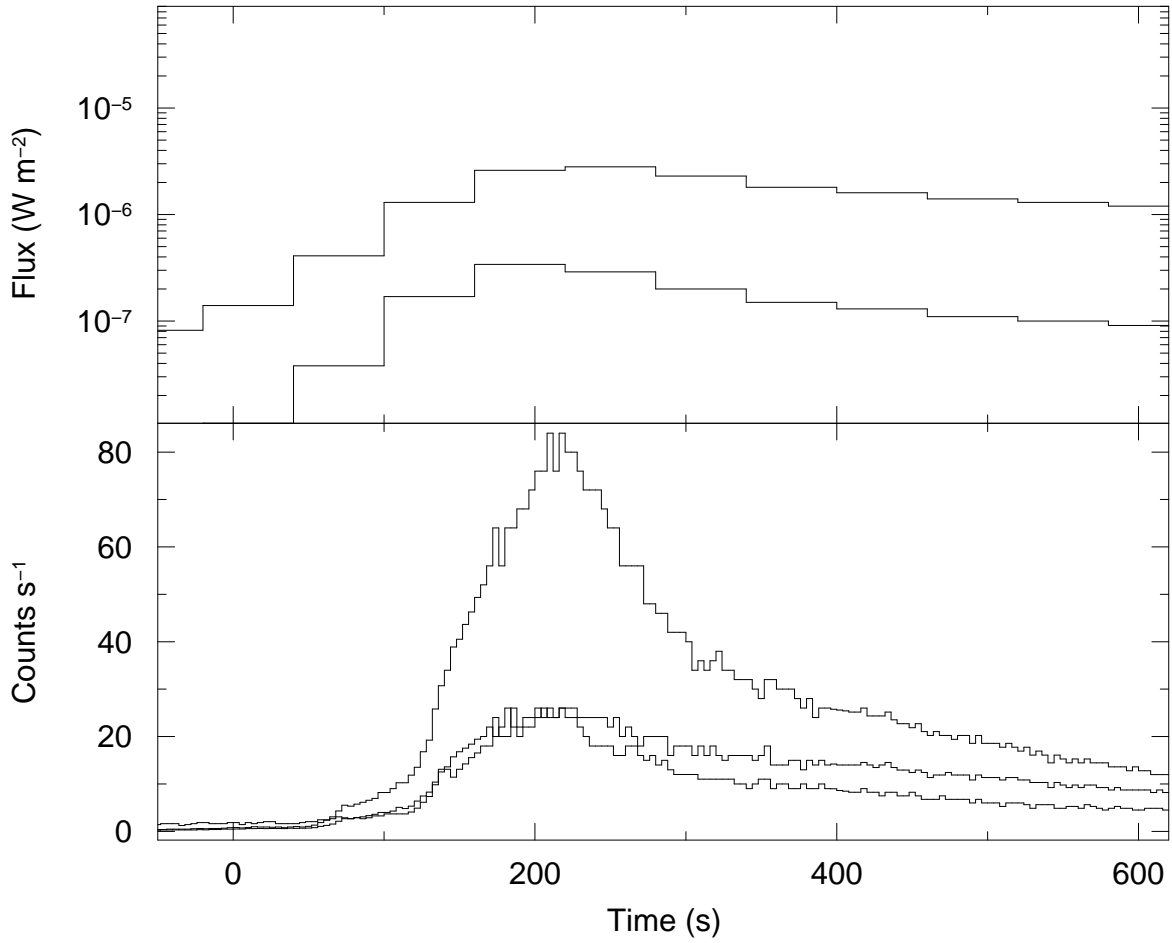}
\caption{  Top panel: GOES light curves in the bands
1 -- 8 $\AA$ (1.6 -- 12.4 keV) and  0.5 -- 4 $\AA$ (3.1- 24.8 keV) 
with the time resolution of 1 minute.
Bottom panel: RHESSI light curves in the
energy bands - 6 -- 12 keV, 3 -- 6 keV  and
12 -- 25 keV (from top). Bin size is 4 seconds. T$_0$ is UT 07:08:50 on 2009 July 5.}
\end{figure*}




\begin{figure*}
\includegraphics[angle=-90,scale=.70]{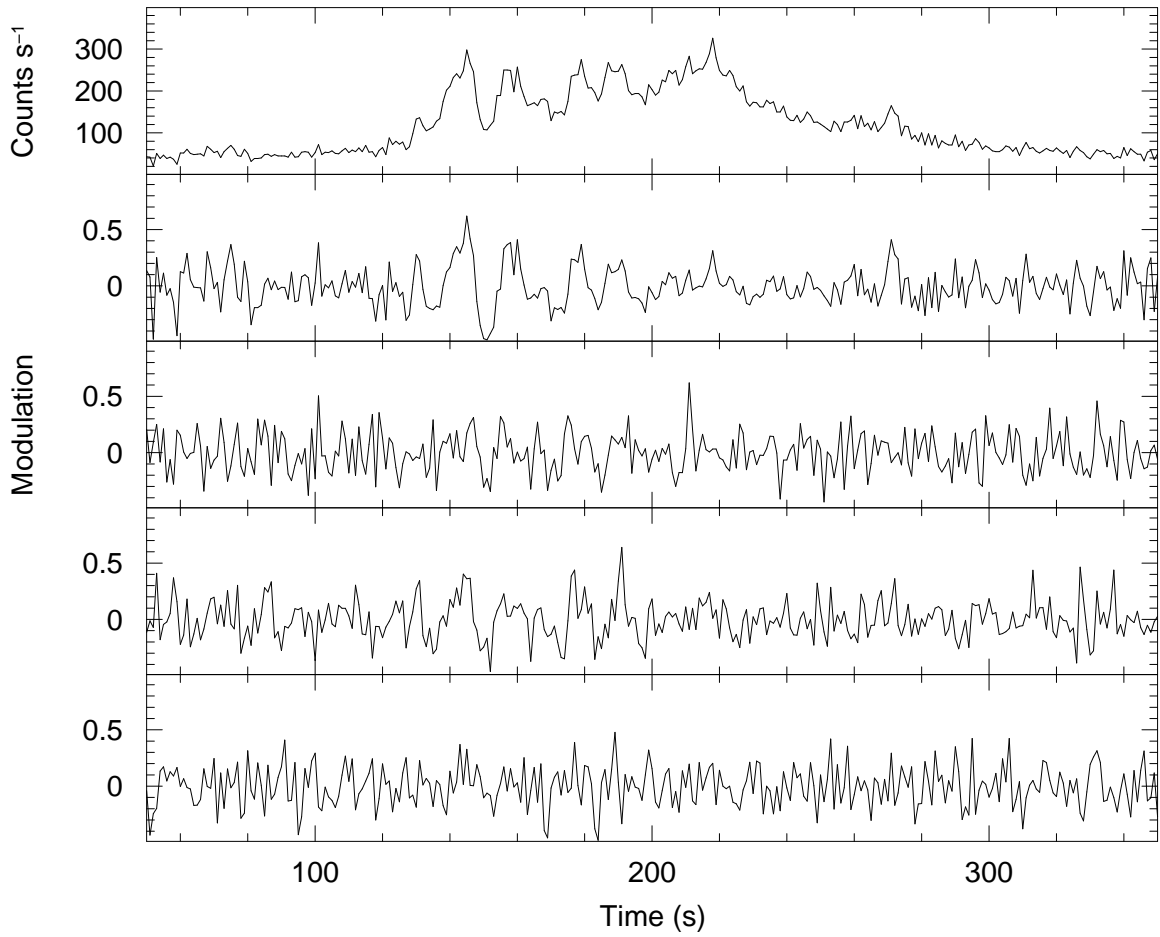}
\caption{ The RT-2/S low energy light curve (top panel) is shown with the
normalized modulation (see text) in the four energy bands shown in Figure 2. 
}
\end{figure*}

\begin{figure}
\includegraphics[angle=-90,scale=.70]{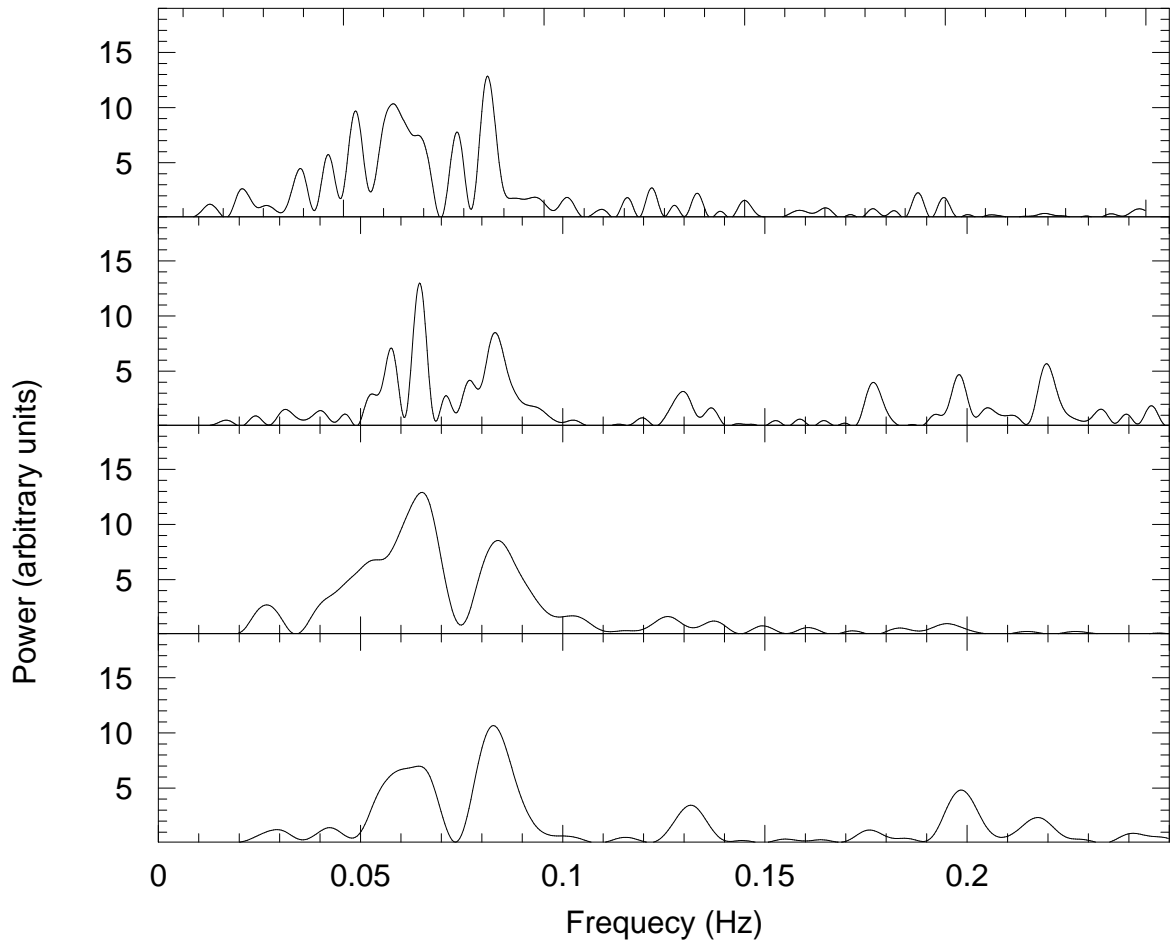}
\caption{The power spectra of the RT-2/S low energy light curves for the
rising phase (top panel) and the falling phase (second panel from the top). The
next two panels show the corresponding power spectra of the RT-2/G low energy light curves. }
\label{fig7}
\end{figure}

\begin{figure}
\includegraphics[angle=-90,scale=.70]{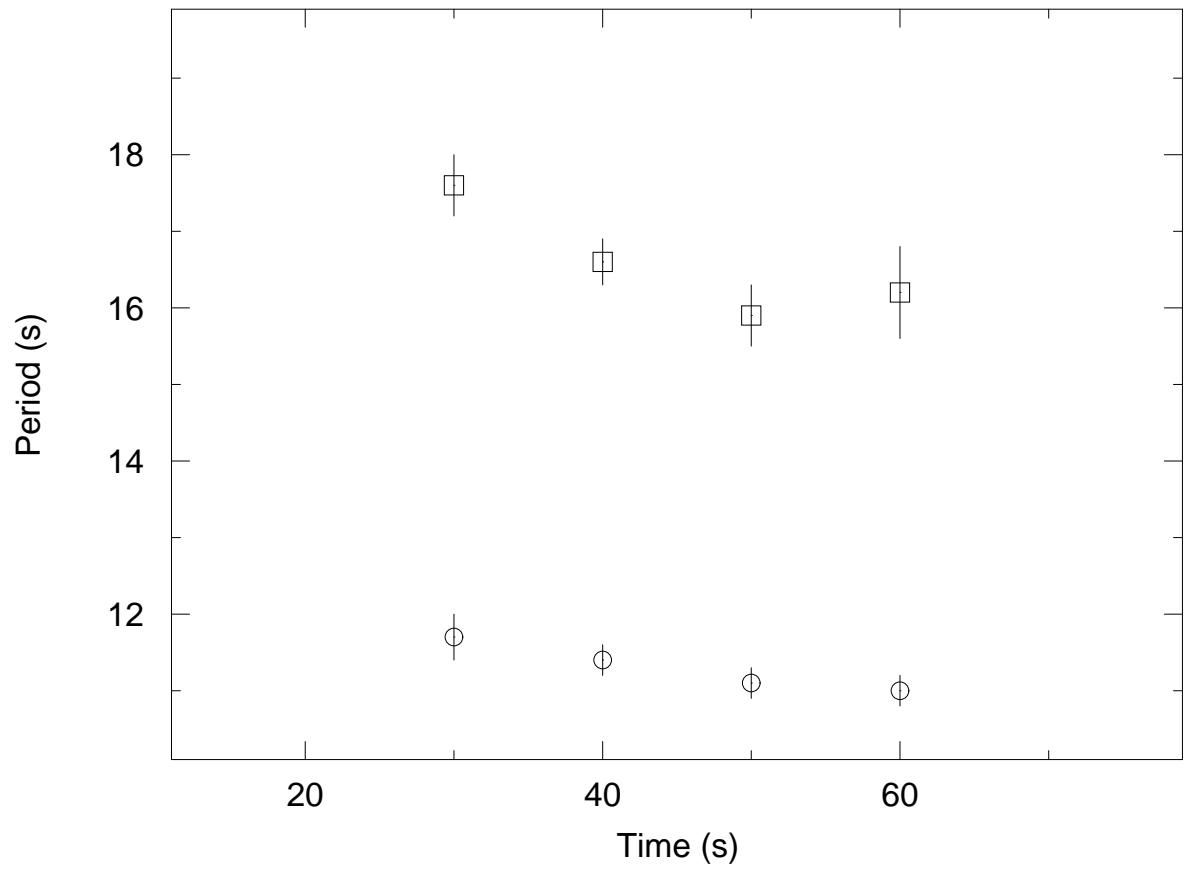}
\caption{The variation of the period of QPPS with time from the start of the flare 
(07:10:55 UT on 2009 July 5)}
\label{fig8a}
\end{figure}

\begin{figure}
\includegraphics[angle=-90,scale=.70]{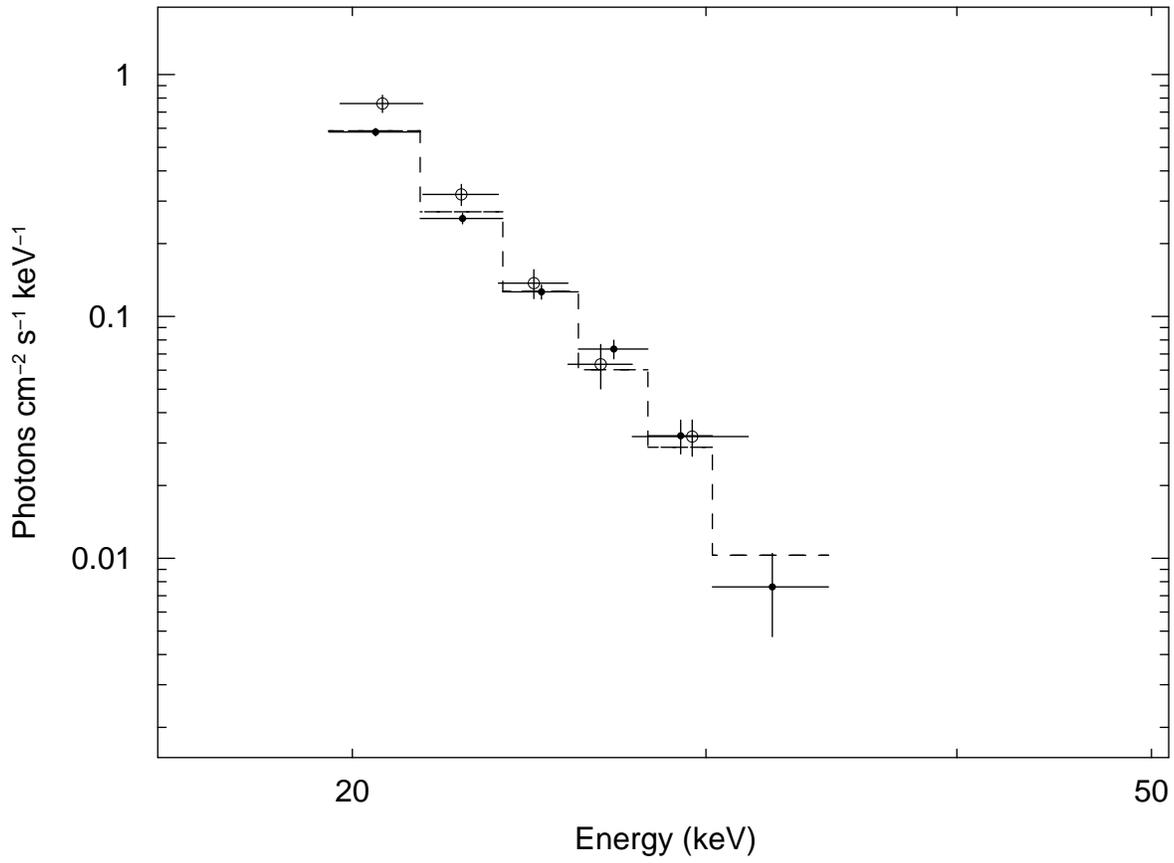}
\caption{The deconvolved spectra from RT-2/S (filled circles) and RT-2/G
(open circles) along with a simple bremsstrahlung spectrum (dashed line).
}
\label{fig8a}
\end{figure}

\begin{figure}
\includegraphics[scale=.70]{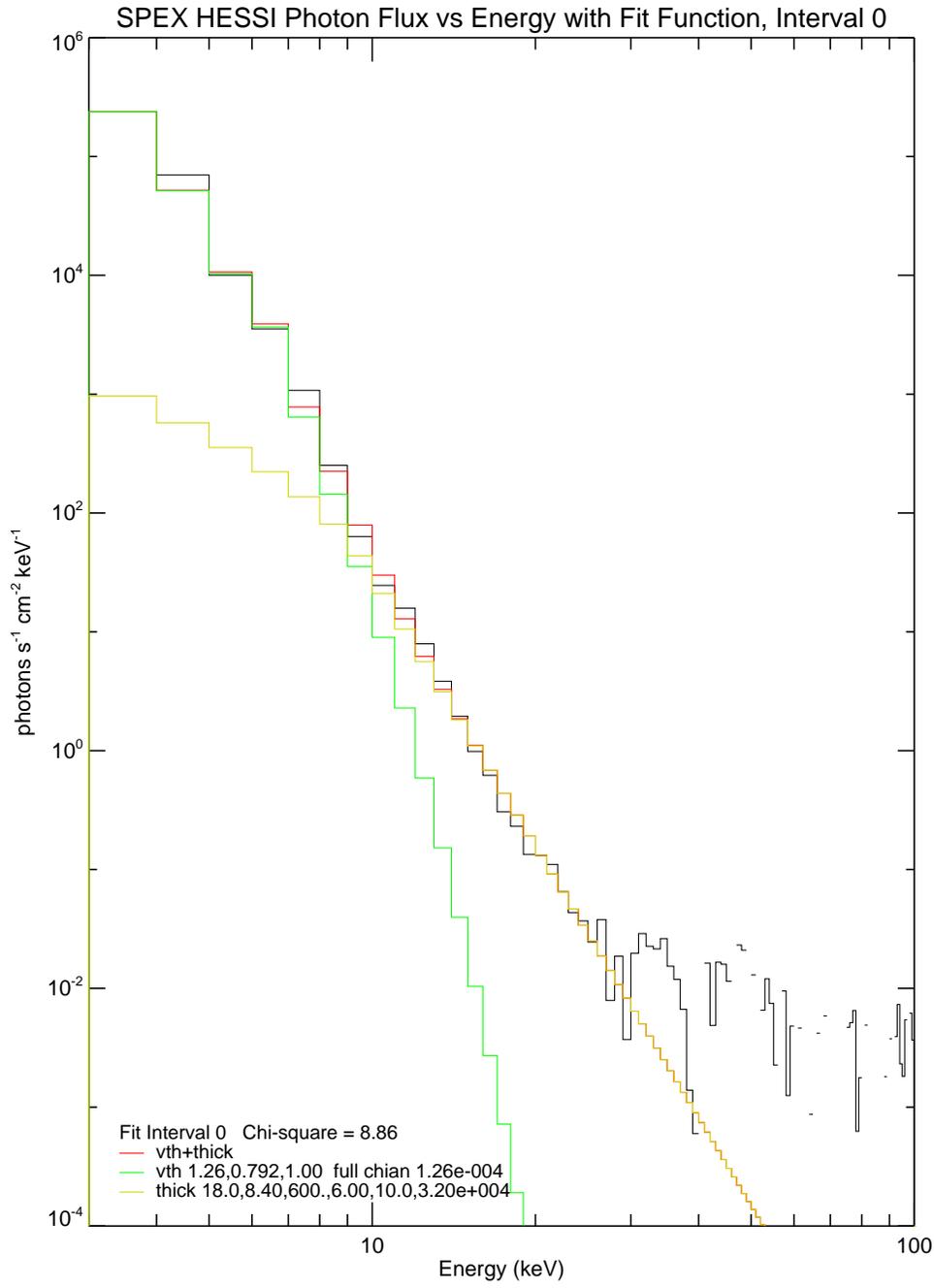}
\caption{The RHESSI photon energy spectra during the 
rising phase of the flare (07:11:40 to 07:12:30). The
two component thin and thick target bremsstrahlung model
is also shown in the figure.}
\label{fig2}
\end{figure}





\begin{table}[h]
\renewcommand{\arraystretch}{1.3}
\caption{Modulation amplitude of the 2009 July 5 flare (in percent) }
\label{First Table}
\centering
\begin{tabular}{ccc}
\hline
 & Rising Phase & Falling Phase \\
\hline
RT-2/S 20 -- 35 keV & 13.5$\pm$0.4 & 5.2$\pm$0.4 \\
$~~~~~~~$35 -- 59 keV & 4.6$\pm$2.8 & $<$ 5.6 \\
RT-2/G 25 -- 35 keV & 6.6$\pm$0.4 & $<$ 0.8 \\
RHESSI 3 -- 6 keV & 7.6$\pm$0.4 & 2.4$\pm$0.4\\
$~~~~~~~$6 -- 12 keV & 6.2$\pm$0.4 & 1.5$\pm$0.4\\
$~~~~~~~$12 -- 25 keV & 5.5$\pm$0.3 & $<$ 0.5 \\
\hline
\end{tabular}
\end{table}

\begin{table}[h]
\renewcommand{\arraystretch}{1.3}
\caption{Quasi-Periodic Pulsation Periods in seconds (with false alarm probability) }
\label{Second Table}
\centering
\begin{tabular}{cccc}
\hline
 & Full flare & Rising Phase & Falling Phase \\
\hline
RT-2/S 20 -- 35 keV & 11.6$\pm$0.1 (7 10$^{-4}$) & 15.3$\pm$0.1 (3 10$^{-4}$) & 12.2$\pm$0.2 (2 10$^{-2}$) \\
 (2$^{nd}$ peak) & -- & 11.9$\pm$0.2 (2 10$^{-2}$) & -- \\
RT-2/G 25 -- 35 keV & 15.5$\pm$0.1 (6 10$^{-4}$) & 12.1$\pm$0.2 (3 10$^{-3}$)  & 15.6$\pm$0.3 (0.23)\\
(2$^{nd}$ peak) & -- & 15.5$\pm$0.3 (0.1) & --\\
Full data & 11.62$\pm$0.6 (2 10$^{-4}$) & 15.5$\pm$0.2 (5 10$^{-4}$) & 12.2$\pm$0.2 (2 10$^{-2}$) \\
(2$^{nd}$ peak) & 15.7$\pm$0.1 (3 10$^{-3}$)  & 11.9$\pm$0.2 (5 10$^{-3}$)  & --\\
\hline
\end{tabular}
\end{table}

\begin{table}[h]
\renewcommand{\arraystretch}{1.3}
\caption{Spectral parameters derived from the RHESSI data for the 2009 July 5 flare }
\label{Third Table}
\centering
\begin{tabular}{ccc}
\hline
 & Rising Phase &  Falling Phase \\
\hline
Time period &  7:11:40 to 07:12:30 & 07:12:35 to 07:14:15 \\
Model: Thin target: & & \\
$~~~~~$ EM (10$^{49}$ cm$^{-3}$) & 1.26 &  0.77 \\
$~~~~~$ kT(keV) & 0.79 & 0.80 \\
Model: Thick target: & & \\
$~~~~~$ electron flux (10$^{35}$ s$^{-1}$) & 18.4 & 11.3 \\
$~~~~~$ $\Gamma$ & 8.4 & 8.6\\
\hline
\end{tabular}
\end{table}

\end{document}